\documentclass[aps,prl,twocolumn]{revtex4}
\usepackage{amssymb}

\usepackage{graphicx}

\begin{document}

\title{Cuprate Fermi orbits and Fermi arcs: 
the effect of short-range antiferromagnetic order}
\author{N.~Harrison, R.D.~McDonald, and J.~Singleton}
\affiliation{National High Magnetic 
Field Laboratory, Los Alamos National Laboratory, MS E536,
Los Alamos, New Mexico 87545}
\date{\today}

\begin{abstract}
We consider the effect of a short antiferromagnetic correlation length
$\xi$ on the electronic bandstructure of the underdoped cuprates.
Starting with a Fermi-surface topology 
consistent with that detected in 
magnetic-quantum-oscillation experiments,
we show that a reduced $\xi$ gives an asymmetric
broadening of the quasiparticle dispersion, resulting in
simulated ARPES data very similar to those
observed in experiment. Predicted features include the presence of
`Fermi arcs' close to $a{\bf k}=(\pi/2,\pi/2)$, where $a$
is the in-plane lattice parameter, without the
need to invoke a $d$-wave pseudogap order parameter. 
The statistical variation in the $k$-space areas of 
the reconstructed Fermi-surface pockets causes the 
quantum oscillations to be strongly damped, even in 
very strong magnetic fields, in agreement with experiment.
\end{abstract}

\maketitle
Magnetic quantum oscillations such as the de Haas-van Alphen and
Shubnikov-de Haas effects are acknowledged to be the 
most reliable method to determine the detailed
low-temperature Fermi-surface topologies of 
metals~\cite{shoenberg1,ref2}.
Recently, Shubnikov-de Haas oscillations have been observed
in the underdoped cuprate superconductors
YBa$_2$Cu$_3$O$_{6.5}$~\cite{doiron1} and 
YBa$_2$Cu$_4$O$_8$~\cite{yelland1,bang};
the data reveal quasi-two-dimensional
Fermi-surface sections with cross-sections that are
$\sim 2-2.4\%$ of the area of the Brillouin zone.
This result is consistent with a picture~\cite{doiron1} in which
antiferromagnetic order breaks translational symmetry,
reconstructing the large Fermi surface
seen in overdoped cuprates~\cite{hussey} into
four small, symmetry-related pockets.
At first sight, this conclusion is at variance with
angle-resolved photo-emission (ARPES) experiments~\cite{shen1,kanigel1},
which find only so-called ``Fermi arcs''.
However, we show here that the quantum-oscillation
and ARPES results are compatible.
Starting with a Fermi surface 
consistent with the quantum-oscillation experiments,
we show that a reduced antiferromagnetic 
correlation length $\xi$ gives an asymmetric
broadening of the quasiparticle dispersion, resulting in
predicted ARPES spectral weights similar to those
observed in experiment.

The breaking of translational symmetry due to antiferromagnetism
can be described by a modulation vector ${\bf Q}$ that is sharply defined 
in an ideal antiferromagnet~\cite{gruner1};
this is the case in the parent Mott-insulating 
phase of the cuprates~\cite{kampf1},
realized for hole dopings $p\lesssim$~0.05.
However, in the underdoped regime (0.05~$\lesssim p\lesssim$~0.2),
slow antiferromagnetic fluctuations, 
detected in neutron-scattering experiments,
cause ${\bf Q}$ to become subject to statistical variations
over a wide range of conditions~\cite{kampf1,birgeneau1}.
At temperatures $T\lesssim$~10~K, the fluctuations become sufficiently slow 
for the system to resemble a glassy state comprising 
antiferromagnetic microdomains with lifetimes 
$\tau_{\rm f}\approx$~10$^{-9}$~s~\cite{hayden1}. 
This lifetime exceeds by several orders of magnitude
both that for the re-equilibration of the Fermi sea occurring 
in ARPES experiments ($\tau_{\rm r}\sim$~10$^{-15}$~s)~\cite{shen1,kanigel1}
and the period of quasiparticle Fermi-surface orbits in 
strong magnetic fields ($2\pi\omega^{-1}_{\rm c}\sim$~10$^{-12}$~s,
where $\omega_{\rm c}$ is the cyclotron frequency~\cite{doiron1,yelland1}). 
Therefore, from the perspective of both experimental techniques,
quasiparticles can be considered to propagate through a glassy 
antiferromagnetic medium in which the staggered magnetization 
is quasistatic and periodic over distances of order $\xi$. 

\begin{figure}[htbp!]
\centering
\vspace{-8mm}
\includegraphics[width=0.45\textwidth]{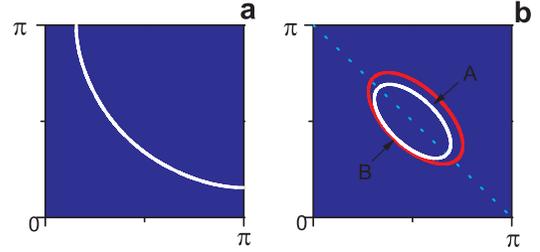}
\vspace{-7mm}
\caption{Quadrant of a notional cuprate Fermi surface 
for $p=$~0.095 doped holes per planar Cu atom. 
{\bf a.}~The Fermi surface 
given by Eq.~\ref{original} in the absence of ordering. 
{\bf b.}~The Fermi surface (white line) after reconstruction 
({\it i.e.} folding about the dotted line) with respect 
to $a{\bf Q}=a{\bf Q}_0=(\pi,\pi)$ with $\Delta=t_{10}$ 
in Eq.~\ref{gapped}. The red line shows the 
corresponding Fermi surface for a slightly different 
${\bf Q}$, shifted relative to ${\bf Q}_0$.} 
\label{Fermisurface}
\vspace{-3mm}
\end{figure}

To see how ARPES and quantum oscillation 
experiments are affected by the short antiferromagnetic correlation
lengths $\xi$, we consider a simplified, two-dimensional quasiparticle
dispersion of the form
\begin{eqnarray}\label{original}
\varepsilon_{\bf k}=2t_{10}(\cos ak_x+\cos ak_y) +\nonumber~~~~~~~~~~~~~~~~~~~~~ 
\\~~~2t_{11}(\cos a(k_x+k_y)+\cos a(k_x-k_y)).
\end{eqnarray}
Eq.~\ref{original} gives rise to a large hole Fermi surface sheet
centered on $a{\bf k}=(\pi,\pi)$ for $t_{10}<0$, 
$t_{11}\approx~-$0.34$\times t_{10}$ and $p>0$;
this is approximately representative of cuprate Fermi surfaces 
in the absence
of ordering~\cite{hussey}.
Figure~\ref{Fermisurface}a shows a quadrant of such a 
Fermi surface for $p=0.095$ holes per Cu atom within the 
Cu-O planes. The entire Fermi surface accommodates $1+p$ 
holes per unit cell (in the absence of ordering), 
having a $k$-space cross-sectional area 
$A_{\bf k}=2\pi^2(1+p)/a^2$, where $a$ is the in-plane lattice
constant. The Fermi energy $\varepsilon_{\rm F}$ 
is obtained self consistently from 
$A_{\bf k}=\int\int{\rm f}(\varepsilon_{\rm F}-\varepsilon_{\bf k}){\rm d}k_x{\rm d}k_y$, 
where ${\rm f}(x)$ is the Fermi-Dirac distribution 
function with the double integration being performed 
within the first Brillouin zone.

The reconstruction of such a Fermi surface caused by long-range 
antiferromagnetic order is well understood~\cite{gruner1}. 
The staggered magnetization ${\bf m}={\bf m}_0\cos({\bf r}\cdot{\bf Q})$ 
represents a new periodicity of the system that
introduces a new, smaller Brillouin zone and causes
the quasiparticle dispersion relationship
(Eq.~\ref{original}) to become gapped close to the
zone boundaries by an energy $2\Delta$. 
In the weak-coupling limit, $\Delta\ll4t_{10}$, 
the modified dispersion can be written
\begin{equation}\label{gapped}
\varepsilon_{{\bf k},{\bf Q}}\approx\frac{\varepsilon_{\bf k}+\varepsilon_{{\bf k}+{\bf Q}}}{2}\pm\sqrt{\big(\frac{\varepsilon_{\bf k}-\varepsilon_{{\bf k}+{\bf Q}}}{2}\big)^2+\Delta^2},
\end{equation}
in the absence of an applied magnetic field~\cite{gruner1,magneticfield}. 
The white line in Fig.~\ref{Fermisurface}b shows one of the 
four equivalent Fermi-surface pockets that result
for $\Delta=t_{10}$ and 
$a|{\bf Q}|=a|{\bf Q}_0|=|(\pi,\pi)|$~\cite{Q0note}. 
To accommodate $p$ holes within the reduced Brillouin zone
(bounded by the dotted line), each pocket has an area 
$A_{{\bf k},{\bf Q}_0}=\pi^2p/a^2$~\cite{ninini}. 
To conserve charge neutrality,
the introduction of antiferromagnetic order
is accompanied by a shift in Fermi energy from $\varepsilon_{\rm F}$ to 
$\varepsilon_{{\rm F},{\bf Q}_0}$;
hence $A_{{\bf k},{\bf Q}_0}=
\frac{1}{4}\int\int{\rm f}(\varepsilon_{{\rm F},{\bf Q}_0}-
\varepsilon_{{\bf k},{\bf Q}_0}){\rm d}k_x{\rm d}k_y$.

The parameters $p=0.095$ and $\Delta=t_{10}$ 
used in Fig.~\ref{Fermisurface}b are chosen to 
mimic the experimental Fermi surfaces of 
Ca$_{1.9}$Na$_{0.1}$CuO$_2$Cl$_2$~\cite{shen1} and 
YBa$_2$Cu$_4$O$_8$~\cite{yelland1} for which the corresponding 
quantum-oscillation frequency 
is $F=hp/4ea^2\approx$~650~T.
Whereas the Mott insulating parent phase ($p\lesssim$~0.05)
may be understood within a strong-coupling description of 
antiferromagnetism~\cite{kampf1,trugman1}, 
the appropriate regime for $p\approx$~0.1 remains a matter of 
conjecture~\cite{birgeneau1}. 
In the strong coupling limit ($\Delta\gg t_{10}$), 
the hole pockets at $(\pi/2,\pi/2)$ would depart only 
marginally from the form depicted in Fig.~\ref{Fermisurface}b, 
the primary effect being to renormalize the widths of the upper and 
lower bands defined by Eq.~\ref{gapped}. 
In the very weak coupling limit ($\Delta/t_{10}\rightarrow$~0), 
however, in addition to the hole pockets at $(\pi/2,\pi/2)$,
small electron pockets would appear at $(\pi,0)$;
there is as yet no evidence for the latter pockets in
quantum-oscillation data~\cite{yelland1}. 

We model the effect of short correlation lengths
using a probability distribution for ${\bf Q}$~\cite{pbl}, 
as determined by neutron-scattering experiments~\cite{kampf1}. 
As stated above, in ARPES and quantum-oscillation experiments
we are in the limit 
$\tau_{\rm f}\gg\omega^{-1}_{\rm c}\gg\tau_{\rm r}$, so that
the distribution can be considered static:
\begin{equation}\label{probability}
p_{\bf Q}=\frac{K}{\xi^{-2}+({\bf Q}_0-{\bf Q})^2},
\end{equation}
with $\int\int p_{\bf Q}{\rm d}Q_x{\rm d}Q_y=1$~\cite{lorentzian}. 
To help understand how the Fermi surface is 
modified by variations in ${\bf Q}$, the red curve in 
Fig.~\ref{Fermisurface}b shows how the pocket is deformed 
by a small shift of ${\bf Q}$ relative to 
${\bf Q}_0$. Whereas the `A' branch of the Fermi-surface orbit
(required by the translational symmetry of the new Brillouin zones) 
is strongly modified by this shift, the `B' branch that corresponds 
more closely to the original Fermi surface 
is less altered. This is true irrespective of the 
direction of the shift in ${\bf Q}$. 
Thus, the `B' branch is more resilient against 
statistical variations in ${\bf Q}$ than is `A'. 
On considering all possible values of ${\bf Q}$ within this distribution, 
the resulting dispersion and Fermi surface become broadened;
observables such as the photoemission 
spectral weight and Landau-level broadening can 
then be estimated by analogy with other types 
of ``phase smearing''~\cite{shoenberg1,alloy}.

ARPES provides information on 
$\varepsilon_{\bf k}$ and the extent to 
which it is broadened by correlations or other factors, giving rise 
to a photoelectron intensity that 
is distributed in energy and $k$-space~\cite{shen1,kanigel1,plate1}. 
When the energy distribution 
curve is cut off by the Fermi-Dirac distribution, the rapidity with 
which the integrated intensity grows 
with $\varepsilon_{\rm F}-\varepsilon_{\bf k}$ close 
to the Fermi energy can be used to obtain 
a `Fermi surface intensity plot'~\cite{shen1,kanigel1,plate1}. 
The discontinuous jump at $\varepsilon_{\rm F}$ encountered 
in conventional metals 
therefore gives rise to an associated spectral weight 
$S_{\bf k}=\delta(\varepsilon_{\bf k}-\varepsilon_{\rm F})$ 
that is infinite at $T=$~0 under ideal conditions. 
In actual experiments, however, with 
overdoped Tl$_2$Ba$_2$CuO$_{6+\delta}$ providing a good example, the 
spectral weight at $\varepsilon_{\rm F}$ is broadened by finite 
temperature and self-energy effects and by instrumental 
energy resolution~\cite{plate1}. 

\begin{figure}[htbp!]
\centering
\includegraphics[width=0.40\textwidth]{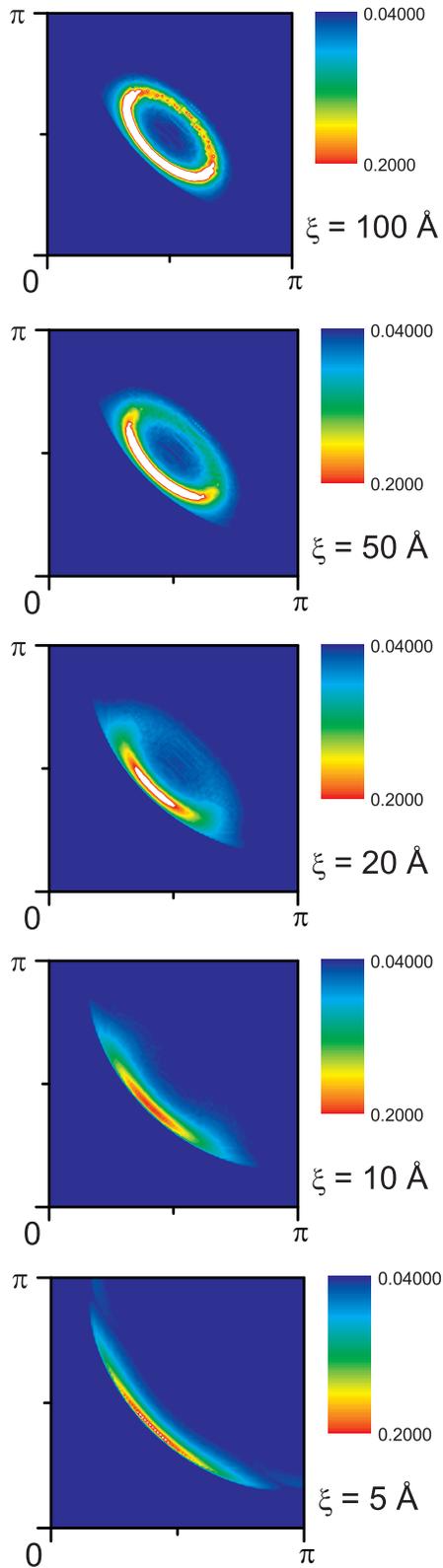}
\caption{Spectral weight calculated over a quadrant of the Fermi 
surface using Eq.~\ref{spectralweight} for different values of 
$\xi$. Since the continuous variables ${\bf k}$ and $\varepsilon$ are 
replaced by discrete quantities in the simulations, the delta 
function is replaced by a Kronecker delta.} 
\label{weight}
\end{figure}

In the case of antiferromagnetism
with a distribution of {\bf Q}, 
we determine the spectral weight from the 
convolution of 
$\delta(\varepsilon_{{\bf k}, {\bf Q}}-
\varepsilon_{{\rm F},{\bf Q}_0})$ with Eq.~\ref{probability}:
\begin{equation}
\label{spectralweight}
S_{{\bf k},{\bf Q}_0}=\int\int
\delta(\varepsilon_{{\bf k},{\bf Q}}-\varepsilon_{{\rm F},{\bf Q}_0})
p_{\bf Q}
{\rm d}Q_x{\rm d}Q_y.
\end{equation}
The sensitivity of the reconstructed Fermi 
surface to statistical variations in ${\bf Q}$
(i.e. Fig.~\ref{Fermisurface}b)
introduces a strongly $k$-space dependent 
broadening of the spectral weight. 
Fig.~\ref{weight} shows 
the spectral weight calculated using Eq.~\ref{spectralweight} 
for several different values of $\xi$. 
At large $\xi$, the spectral weight strongly resembles the 
outline of the Fermi surface depicted in white in 
Fig.~\ref{Fermisurface}b. 
As $\xi$ becomes shorter, however, the spectral weight
corresponding to the part of the Fermi surface 
produced by the $k$-space reconstruction (labeled `A')
fades to be replaced 
by a greater concentration along an arc of the 
original unreconstructed Fermi surface. 
At $\xi=$~10 and 20~\AA, the 
distribution of spectral weight closely resembles 
that observed in Ca$_{1.9}$Na$_{0.1}$CuO$_2$Cl$_2$
(see Figs.~1b,c of Ref.~\cite{shen1}), for which $p\approx 0.1$. 
These values of $\xi$ are in excellent agreement 
with neutron-scattering estimates made at the same nominal $p$ in a 
variety of cuprates (at temperatures $\sim$ tens of kelvin in 
zero applied magnetic field)~\cite{kampf1}.

\begin{figure}[htbp!]
\centering
\includegraphics[width=0.43\textwidth]{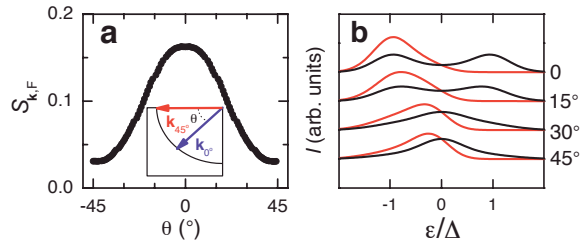}
\vspace{-4mm}
\caption{{\bf a} The approximately sinusoidal 
variation of the spectral weight in Fig.~\ref{weight} ($\xi=$~10~\AA) 
with angle $\theta$, following the arc of the 
unreconstructed Fermi surface from Fig.~\ref{Fermisurface}a. $\theta$ 
is the angle between the line linking $(\pi,\pi)$ to $(0,0)$ 
and that linking $(\pi,\pi)$ to the arc (inset). 
{\bf b}~Calculated energy distribution intensity 
at ${\bf k}_{\rm F}$ along the unreconstructed Fermi 
surface at different $\theta$. Red curves are
calculations; black curves are ``symmetrized''
in the same manner as experimental data~\cite{kanigel1}.
} 
\label{intensity}
\end{figure}

The model reproduces the features seen 
in Ca$_{1.9}$Na$_{0.1}$CuO$_2$Cl$_2$ and 
Bi$_2$Sr$_2$CaCu$_2$O$_{8+\delta}$ ARPES 
data~\cite{shen1,kanigel1}.
Fig.~\ref{intensity}a shows the 
angle ($\theta$) dependence of the spectral weight 
from Fig.~\ref{weight} ($\xi=10~$\AA) tracing the 
${\bf k}={\bf k}_ {\rm F}$ arc of the original 
unreconstructed Fermi surface in Fig.~\ref{Fermisurface}a. 
The sinusoidal variation of 
$S_{{\bf k}_{\rm F},{\bf Q}_0}$ with 
$\theta$ (Fig.~\ref{intensity}a) 
and the strongly reduced gradient of 
the energy distribution intensity $I$
with respect to $\varepsilon$ at ${\bf k}_{\pm 45^\circ}$
in Fig.~\ref{intensity}b
are very similar to ARPES data~\cite{shen1,kanigel1}.
Such features have been interpreted in
terms of a `pseudogap';
note that our model
shows that short-range antiferromagnetic order 
can also give the `illusion' of a 
{\it d}-wave order parameter 
with a node at $\theta=~0^\circ$ and antinodes at 
$\theta=\pm45^\circ$ (see inset)~\cite{shen1}. 
This is also shown in 
Fig.~\ref{intensity}b that plots the energy distribution 
calculated using Eq.~\ref{spectralweight} at selected values 
${\bf k}_{\rm F}$ along the arc of the unreconstructed
Fermi surface. To simulate experimental curves, the 
spectral intensity is modulated by 
$f(\varepsilon_{\bf k}-\varepsilon_{\rm F})$ at $T=$~0 
and broadened by an `instrumental resolution' 
gaussian of width $\approx\frac{1}{5}\Delta$. 
When symmetrized~\cite{kanigel1}, the energy distribution 
closely reproduces the apparent ``pseudogap behavior'' of 
Bi$_2$Sr$_2$CaCu$_2$O$_{8+\delta}$~\cite{kanigel1}.

\begin{figure}
\centering
\includegraphics[width=0.45\textwidth]{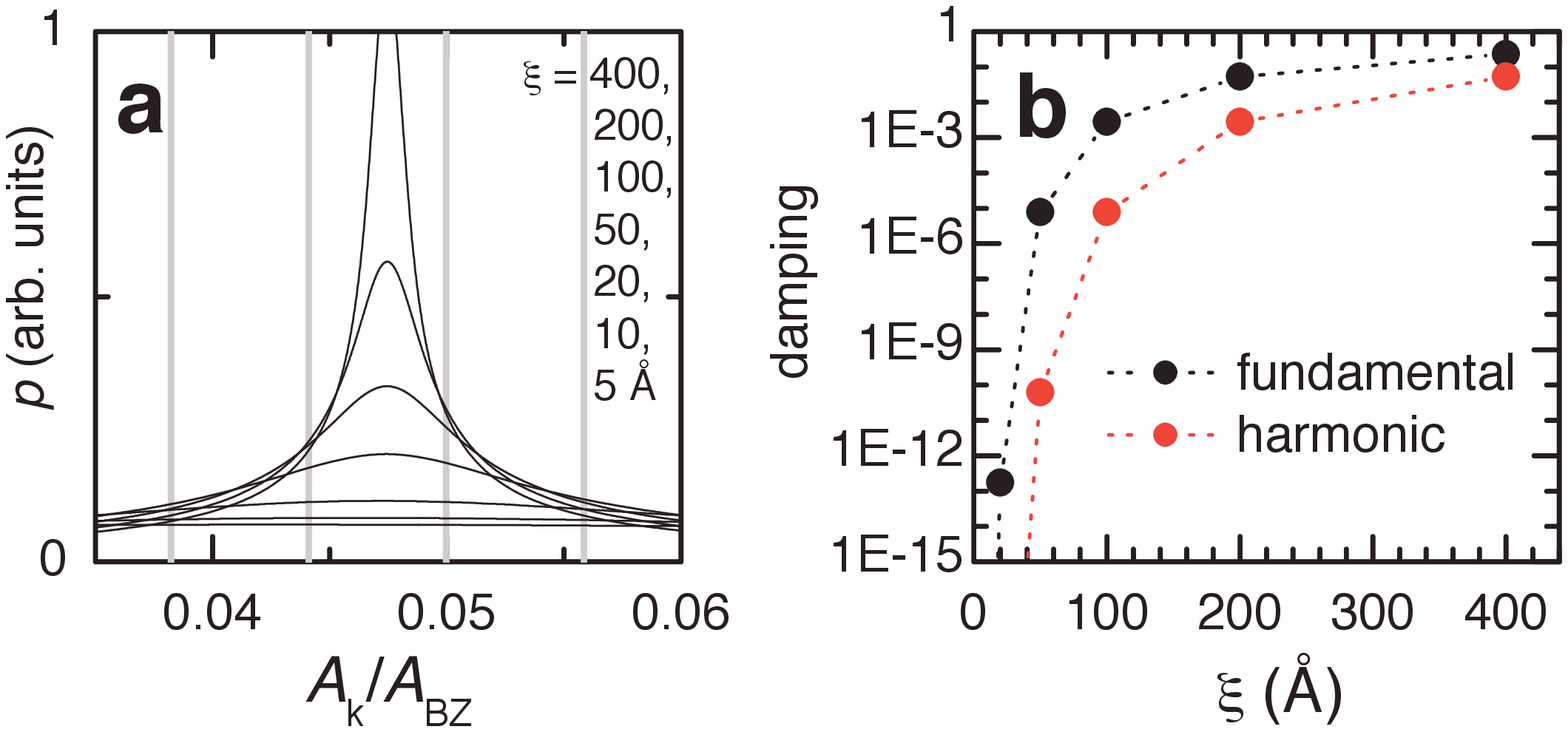}
\vspace{-5mm}
\caption{{\bf a} The $A_{{\bf k},{\bf Q}_0}$ probability 
distribution for several values of $\xi$. 
Grey lines mark the positions of the ideal $\delta$-function 
Landau levels for a field of 80~T
expected in this 
quasi-two dimensional metal were long-range antiferromagnetic 
order established. These would furnish quantum 
oscillations periodic in $1/B$, rich in harmonic content. 
{\bf b} A plot of the corresponding damping factor 
versus $\xi$ for the fundamental frequency (black) and 
harmonic (red) resulting from the convolution of the 
probability distribution with the $\delta$-function Landau levels.} 
\label{quantum}
\vspace{-4mm}
\end{figure}

Quantum-oscillation experiments
give a direct measure of $A_{{\bf k},{\bf Q}_0}$ 
(Fig.~\ref{Fermisurface}b) that is also subject to 
statistical variations. The probability distribution is given by 
\begin{equation}
\label{areaprobability}
p_{A_{{\bf k}}}=\int\int
\delta(A_{{\bf k},{\bf Q}}-A_{{\bf k}})p_{\bf Q}
{\rm d}Q_x{\rm d}Q_y.
\end{equation}
Fig.~\ref{quantum}a shows Eq.~\ref{areaprobability}
for several $\xi$; the curves resemble the probability 
distributions caused by finite quasiparticle lifetime 
or other forms of disorder~\cite{shoenberg1,alloy}, 
leading to a qualitatively similar Landau-level
broadening and quantum-oscillation damping. 
Figure~\ref{quantum}b shows predicted damping factors 
for the fundamental quantum-oscillation frequency
$F\approx$~650~T in YBa$_2$Cu$_4$O$_8$~\cite{yelland1} 
and for the $2F$ harmonic. The correlation lengths 
$\xi\approx$~10 and 20~\AA~ that are best able to reproduce the 
${\bf k}$-dependent spectral weight of 
Ca$_{1.9}$Na$_{0.1}$CuO$_2$Cl$_2$~\cite{shen1} in 
Fig.~\ref{weight} cause too much damping
to mimic the data of Ref.~\cite{yelland1};
a damping factor $\sim$~10$^{-3}$ 
is the typical cutoff for observable quantum 
oscillations, corresponding to $\xi\approx$~90~\AA.
An upper limit of $\xi \lesssim 400$\AA~ comes from the
absence of harmonics in the data,
normally abundant in quasi-two dimensional 
metals~\cite{harrison1}; {\it i.e.} $90 \lesssim\xi\lesssim 400$~\AA. 
The observation of quantum oscillations in 
YBa$_2$Cu$_3$O$_{6.5}$~\cite{doiron1} and 
YBa$_2$Cu$_4$O$_8$~\cite{yelland1} therefore suggests
a correlation length several times longer 
than that in ARPES experiments. 

The differences in $\xi$ result from the contrasting conditions 
under which ARPES and quantum-oscillation experiments are performed. 
Whereas ARPES experiments are performed at tens 
of kelvins and zero magnetic field, quantum-oscillation 
experiments are made at liquid helium temperatures in the 
strongest fields available. 
Abundant neutron scattering 
data on Sr- and O-doped La$_2$CuO$_4$ (as well as electron-doped 
cuprates) show an enhancement of antiferromagnetism 
and/or its associated correlation length at liquid $^4$He 
temperatures under magnetic 
fields~\cite{neutron},
with values as large as $\xi\approx 400$~\AA~ 
reported~\cite{neutron}. 
While similar neutron-scattering data are 
unavailable for the two systems
(YBa$_2$Cu$_3$O$_{6.5}$ and YBa$_2$Cu$_4$O$_8$) in which 
quantum oscillations are observed, 
alternative experimental techniques do provide 
evidence for field-induced antiferromagnetism~\cite{zhang1}. 
These include nuclear magnetic resonance~\cite{mitrovic1}, 
muon spin rotation~\cite{miller1} and magnetic torque~\cite{ishida1}.

In conclusion, we show that the seemingly contradictory 
results of ARPES and quantum oscillation 
experiments on the underdoped cuprates can be brought 
into mutual agreement by considering the
slow antiferromagnetic fluctuations with a short correlation length
detected by neutron-scattering experiments.
Whereas the correlation length is very short (10~$\lesssim\xi\lesssim$~20~\AA) 
in the normal state at zero magnetic field, 
it becomes enhanced (90~$\lesssim\xi\lesssim$~400~\AA) at liquid 
He temperatures in magnetic fields due to a field-induced 
stabilization of antiferromagnetism~\cite{chen}. 
Short correlation lengths lead to a reduction in the spectral 
weight that is most pronounced close to $(\pi,0)$ on 
the unreconstructed Fermi surface, giving rise to the emergence of Fermi 
arcs near $(\pi,\pi)$ in a qualitatively similar fashion to 
a ${\it d}$-wave order parameter. 
The effects described in this paper should also occur 
for other forms of ordering that lack long range periodicity, 
including stripes and/or charge order~\cite{shen1}. 
However, only antiferromagnetism is truly ubiquitous, 
giving rise to clear signatures in neutron-diffraction 
experiments in all underdoped cuprates investigated.

We are very grateful to Ed Yelland, Chandra Varma
and Louis Taillefer
for stimulating discussions.
This work is supported by the
US Department of Energy (DoE) BES
program ``Science in 100~T''.
Work at NHMFL is carried out under
the auspices
of the National Science
Foundation, DoE and the State of Florida.


\begin{thebibliography}{99}
\bibitem{shoenberg1} 
D.~Shoenberg, 
{\it Magnetic Quantum Oscillations in Metals} 
(Cambridge University Press, 1984).
\bibitem{ref2}
A.~Wasserman and M.~Springford, Adv. Phys. {\bf 45}, 471 (1996);
J.~Singleton, Rep. Prog. Phys. {\bf 63}, 1111 (2000).
\bibitem{doiron1} 
N.~Doiron-Leyraud {\it et al.}, Nature {\bf 447}, 565 (2007).
\bibitem{yelland1} 
E.A.~Yelland {\it et al.}, preprint arXiv:0707.0057.
\bibitem{bang}
A.F.~Bangura {\it et al.}, preprint arXiv:0707.4461.
\bibitem{hussey} 
N.E. Hussey et al., Nature {\bf 425}, 814 (2003).
\bibitem{shen1} 
K.~M.~Shen {\it et al.}, Science {\bf 307}, 901 (2005).
\bibitem{kanigel1} 
A.~Kanigel {\it et al.}, Nature Phys. {\bf 2}, 447 (2006).
\bibitem{gruner1} 
G.~Gr\"{u}ner, Rev.  Mod. Phys. {\bf 66}, 1 (1994);
R.G.~Goodrich {\it et al.}, Phys. Rev. Lett. {\bf 97}, 146404 (2006).
\bibitem{kampf1} 
A.~P.~Kampf, Phys. Rep. {\bf 249}, 219 (1994).
\bibitem{birgeneau1} 
R.~J.~ Birgeneau {\it et al.}, J. Phys. Soc. 
Japan {\bf 75}, 111003 (2006).
\bibitem{hayden1} 
S.~M.~Hayden {\it et al.}, Phys. Rev. Lett. {\bf 66}, 821 (1991).
\bibitem{magneticfield} 
The largest field applied (85~T~\cite{yelland1}) is 
significantly lower than that required to polarize the Cu spins~\cite{kampf1}.
\bibitem{Q0note} 
For simplicity we assume $a{\bf Q}_0=(\pi,\pi)$ throughout. 
Whereas the actual ${\bf Q}_0$ may become incommensurate 
for $p\gtrsim$~0.05~\cite{kampf1}, 
branch `B' of the reconstructed Fermi surface 
is just as resilient to small changes in 
${\bf Q}_0$ as it is to small statistical 
variations in ${\bf Q}$. An incommensurate 
${\bf Q}_0$ leads to small differences 
(i.e. see Fig.~\ref{Fermisurface}b) in Fermi surface 
topology for large values of $\xi$ that nevertheless 
become indiscernible for short values of $\xi$.
\bibitem{ninini} 
The estimates of $p$ from quantum-oscillation frequencies made
in \cite{doiron1} and \cite{yelland1} are a factor 2 too large,
through not allowing for the reduced Brillouin zone. See also~\cite{chen}.
\bibitem{chen}
Wei-Qiang Chen {\it et al.}, preprint arXiv:0706.3556.
\bibitem{trugman1} 
S.~A.~Trugman, Phys. Rev. Lett.  {\bf 65}, 500(1990).
\bibitem{pbl}
P.A.~Lee {\it et al.}, Phys. Rev. Lett. {\bf 31},
462 (1973).
\bibitem{lorentzian} 
To avoid logarithmic divergences, 
the $k$-space integration of $p_{\bf Q}$ is 
cutoff at $\pm\pi$. This problem may also 
be avoided by adopting a Gaussian distribution~\cite{kampf1}.
\bibitem{alloy} 
N. Harrison and J. Singleton, J. Phys.: Cond. Matter {\bf 13}, L463 (2001).
\bibitem{plate1} 
M.~Plat\'{e} {\it et al}, Phys. Rev. Lett. {\bf 95}, 077001 (2005).
\bibitem{harrison1} 
N.~Harrison {\it et al.}, Phys. Rev. B {\bf 54} 9977 (1996)
\bibitem{neutron} 
H.~Kimura {\it et al.}. Phys. Rev. B {\bf 59}, 6517 (1999);
B.~Lake {\it et al.}, Nature {\bf 415}, 299 (2002); 
B.~Lake {\it et al.}, Phys. Stat. Sol. B {\bf 241}, 1223 (2004);
B.~Khaykovich {\it et al.}, Phys. Rev. B {\bf 66}, 014528 (2002);
M.~Matsuura {\it at al.}, Phys. Rev. B {\bf 69}, 104510 (2004);
J.~E.~Sonier {\it et al.}, Physica C {\bf 408-410}, 783 (2004);
B.~Khaykovich {\it et al.}, Phys. Rev. B {\bf 71}, 220508 (2005).
\bibitem{zhang1} 
Y.~Zhang, E.~Demler and S.~Sachdev, Phys. Rev. B {\bf 66}, 094501 (2002).
\bibitem{mitrovic1} 
V.F.~Mitrovic {\it et al.}, Nature {\bf 413}, 501 (2001).
\bibitem{miller1} 
R.I.~Miller {\it et al.}, Phys. Rev. Lett. {\bf 88}, 137002 (2002).
\bibitem{ishida1} 
T.~Ishida {\it et al.}, Physica C {\bf 426-431}, 69 (2005).
\end{thebibliography}
\end{document}